# Unidirectional reflection lasing based on destructive interference and Bragg scattering modulation in defective atomic lattice

Xinfu Zheng,[1] Chen Peng,[1] Duanfu Chen,[1] Tinggui Zhang,[2]
Hanxiao Zhang,[1] Dong Yan,[1] Jinhui Wu[†],[3] and Hong Yang[*][1]

[1]*School of Physics and Electronic Engineering, Hainan Normal University, Haikou 571158, People's Republic of China*
[2]*School of Mathematics and Statistics, Hainan Normal University, Haikou 571158, People's Republic of China*
[3]*School of Physics and Center for Quantum Sciences,
Northeast Normal University, Changchun 130024, People's Republic of China*
(Dated: December 30, 2025)

The novel and ingenious scheme we propose for achieving unidirectional reflection lasing (URL) involves integrating a one-dimensional (1D) defective atomic lattice with a coherent gain atomic system. Its physical essence lies in the fact that the right-side reflectivity is drastically reduced due to the destructive interference between primary and secondary reflections, whereas on the left-side primary reflection is effectively suppressed and the secondary reflection is efficiently enhanced, ultimately reaching the lasing threshold. Through numerical results and further analyses, we have elucidated how to precisely tailor the lattice parameters and coupling fields to control destructive interference point (DIP), thereby realizing URL and enabling its active modulation. Our scheme is experimentally feasible and not only effectively circumvents the stringent conditions faced in directly realizing URL, providing a new pathway, but also beneficial for integrating active photonic devices into compact quantum networks and may improve the efficiency of optical information transmission.

## I. INTRODUCTION

In the past decade or so, optical nonreciprocity has garnered extensive research attention. From the irreversible manipulation based on the magneto-optical Faraday effect [1, 2] to the establishment of magnet-free optical non-reciprocal systems [3–15], and further from the perfect nonreciprocity of unidirectional light propagation [16–20] to the realization of unidirectional light amplification [21–25] and unidirectional lasing [26–28], research in this field has been steadily advancing toward integrated, high-performance photonic devices. In particular, magnet-free unidirectional lasing not only protects lasers to ensure their stable operation but also isolates backscattered light [29–33], avoids unnecessary reflections and crosstalk, and safeguards fragile light sources. As the foundation of non-reciprocal photonic networks, it further promotes the development of integrated photonic devices [34–40].

So far, achieving higher precision in chip-integrated all-optical controlled non-reciprocal devices remains a significant challenge. Nevertheless, substantial progress has been made in the research of non-reciprocal lasing across diverse physical systems. For example, the non-reciprocal Brillouin lasing within silicon waveguides [41–44], the non-reciprocal magnon lasing in hybrid cavity optomagnonical system [45–47], the non-reciprocal phonon lasing in coupled optomechanical and nonlinear-optical resonator [48, 49]. However, the unidirectional lasing that can be indispensable components of photonic integrated circuitry, requires more ingenious design and precise control. The unidirectional lasing modes associated with the frozen mode regime of nonreciprocal slow-wave structures have been proposed, which can only exist in cavities with broken time reversal and space inversion symmetries [50, 51]. Stable unidirectional lasing in ring resonators has been investigated by adding an S-shaped waveguide element to the ring resonator to form a so-called "Taiji" resonator (TJR) [52, 53]. Evidently, the combination of nonreciprocity and optical cavities is the core mechanism for achieving nonreciprocal lasing or unidirectional lasing [54, 55]. However, cavity losses induced by factors such as diffraction effects and mode competition, coupled with mode stability issues, represent critical bottlenecks limiting the performance enhancement of cavity-based lasing. Despite continuous technological innovations driving devices to constantly break physical limits in terms of output power and beam quality, developing a novel system capable of replacing traditional optical cavities, while simultaneously embodying distributed feedback functionality and nonreciprocal characteristics remains a research topic of significant academic value and application prospects. This exploration is not only expected to fundamentally overcome the inherent drawbacks of optical cavities but also provide crucial support for the high-performance optimized design of integrated optical circuits.

In this paper, we propose a novel and experimentally feasible scheme for achieving unidirectional reflection lasing (URL) in a one-dimensional (1D) defective atomic lattice (which introduces spatial symmetry breaking and a distributed feedback (DFB) regime, replace the optical cavity) by leveraging the coherent gain atomic system (which provides probe gain through the spontaneous emission effect). We have achieved a unidirectional photonic reflector and unidirectional reflection amplification in the 1D atomic lattice [56, 57]. Incidentally, lasing output has been achieved in perfect optical lattices theoretically and experimentally, respectively [58, 59]. Further-more, we demonstrate that the URL corresponds to the two eigenvalues of scattering matrix $\lambda_{S^{-1}}^+ \simeq \lambda_{S^{-1}}^- \to 0$, originating from the multiple amplification of left reflection enabled by the combination of symmetry breaking in defect-containing atomic lattices, distributed feedback effects, and gain atomic systems, as well as the near-zero right reflection induced by destructive interference of Bragg scattering.

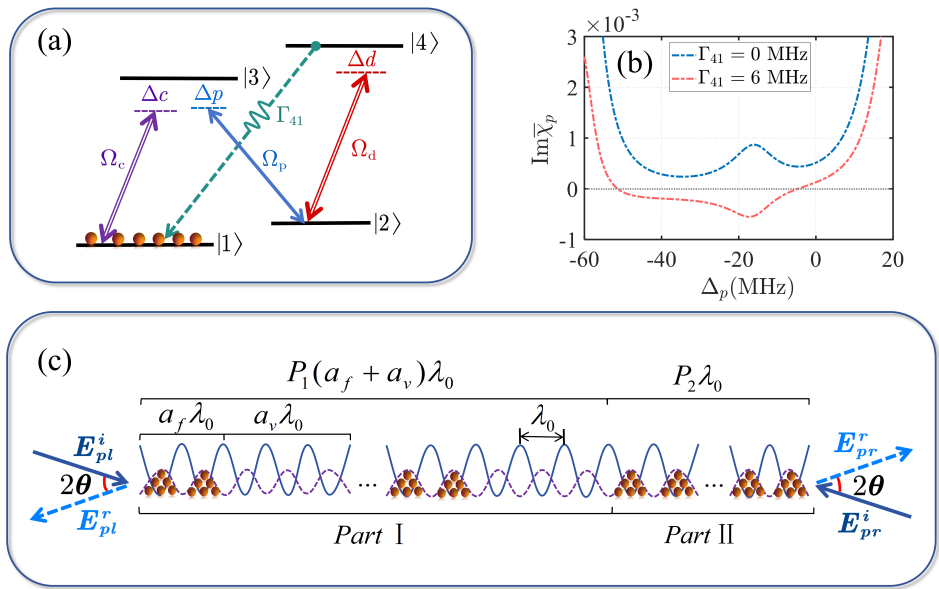

FIG. 1: (a) Energy level diagram of a closed-loop four-level $N$-type atomic system driven by a weak probe field $\Omega_p$, and two strong coupling fields $\Omega_c$ and $\Omega_d$ with the help of spontaneous emission decay rate $\Gamma_{41}$. (b) The imaginary part of average probe susceptibility Im$\overline{\chi}_p$ versus probe detuning $\Delta_p$ with $\Gamma_{41} = 0$ MHz (blue line) and $\Gamma_{41} = 6$ MHz (red line) in $\Delta_c = -15$ MHz, $\Delta_d = 20$ MHz. (c) The 1D defective atomic lattice with the width $\lambda_0$ for each period, which comprises Part I and Part II. The probe filed $\mathbf{E}_p$ is incident from either the left-side (denoted by $\mathbf{E}_{pl}^i$) or the right-side (denoted by $\mathbf{E}_{pr}^i$) with a small incident angle $\theta$ relative to $x$-axis. The relevant reflected fields are denoted by $\mathbf{E}_{pl}^r$ and $\mathbf{E}_{pr}^r$. The parameters are $\Omega_c = 40$ MHz, $\Omega_d = 20$ MHz, $\Omega_p = 0.01$ MHz, $\Gamma_{43} = \Gamma_{21} = 0$ MHz, $\Gamma_{31} = \Gamma_{32} = \Gamma_{41} = \Gamma_{42} = 6$ MHz, $\lambda_{Lat} = 781.3$ nm, $\lambda_{32} = 780.24$ nm, $\Delta\lambda_{Lat} = 0.9$ nm, $\eta = 5.0$, $N_0 = 7 \times 10^{11}$cm$^{-3}$, $\mathbf{d}_{32} = 1.4632 \times 10^{-29}$ C·m.

## II. THEORETICAL MODEL AND EQUATIONS

As shown in Fig. 1(a), the cold $^{87}$Rb atoms are driven into a closed-loop four-level $N$-type system, by a weak probe field with frequency $\omega_p$ (amplitude $\mathbf{E}_p$) and two strong coupling fields with frequencies $\omega_c$ and $\omega_d$ (amplitudes $\mathbf{E}_c$ and $\mathbf{E}_d$) in a defective atomic lattice. The Rabi frequency (detuning) of probe is denoted by $\Omega_p = \mathbf{E}_p \cdot \mathbf{d}_{23}/2\hbar$ ($\Delta_p = \omega_{32} - \omega_p$) interact via the dipole-allowed transition $|2\rangle \leftrightarrow |3\rangle$. The Rabi frequencies (detunings) of two coupling fields $\Omega_c = \mathbf{E}_c \cdot \mathbf{d}_{13}/2\hbar$ ($\Delta_c = \omega_{31} - \omega_c$) and $\Omega_d = \mathbf{E}_d \cdot \mathbf{d}_{24}/2\hbar$ ($\Delta_d = \omega_{42} - \omega_d$) drives the dipole-allowed transition $|1\rangle \leftrightarrow |3\rangle$ and $|2\rangle \leftrightarrow |4\rangle$, respectively. The matrix element $\mathbf{d}_{ij} = \langle i|\mathbf{d}|j\rangle$ is used to denote the dipole moment of transition $|i\rangle$ to $|j\rangle$. $\Gamma_{31}$ and $\Gamma_{32}$ ($\Gamma_{41}$ and $\Gamma_{42}$) denote the spontaneous decay rates the atoms transition from level $|3\rangle$ ($|4\rangle$) to level $|1\rangle$ and $|2\rangle$, respectively. Specifically, the energy levels $|1\rangle$, $|2\rangle$, $|3\rangle$, and $|4\rangle$ refer to $|5S_{1/2}, F = 2, m_F = 1\rangle$, $|5S_{1/2}, F = 2, m_F = 2\rangle$, $|5P_{3/2}, F = 2, m_F = 2\rangle$, and $|5P_{3/2}, F = 3, m_F = 1, 2\rangle$, respectively, in the D2 line of $^{87}$Rb atoms. Noting, the atoms can be transferred from level $|1\rangle$ to level $|3\rangle$ by the coupling field $\mathbf{E}_c$, and spontaneous emission to level $|2\rangle$ emits a photon with the same frequency as probe field $\omega_p$, then transferred to level $|4\rangle$ with the help of coupling field $\mathbf{E}_d$ and spontaneous emission back to level $|1\rangle$. This closed-loop process enables the system to realize the gain of the probe field in population density $N_0 = 7 \times 10^{11}$cm$^{-3}$. In Fig. 1(b), for $\Gamma_{41} = 0$ MHz, the probe susceptibility exhibits an electromagnetically induced transparency (EIT) window over $\Delta_p \in [-50$ MHz, 10 MHz], whereas $\Gamma_{41} = 6$ MHz, this EIT window transforms into a gain window. Nevertheless, the probe gain in the defective lattices remains very low, being only $0.5 \times 10^{-3}$, and the Fig. 1(c) illustrates several structural parameters of a defective lattice.

Within the electric-dipole and rotating-wave approximations, the atom-field interaction Hamiltonian of the system can be expressed as

$$H_I = -\hbar \begin{bmatrix} 0 & 0 & \Omega_c^* & 0 \\ 0 & \Delta p - \Delta c & \Omega_p^* & \Omega_d^* \\ \Omega_c & \Omega_p & -\Delta c & 0 \\ 0 & \Omega_d & 0 & \Delta p - \Delta c - \Delta d \end{bmatrix}. \quad (1)$$

Then, the equations of the density matrix are written as

$$\begin{aligned}
\dot{\rho}_{11} &= i\Omega_c^*\rho_{31} - i\Omega_c\rho_{13} + \Gamma_{21}\rho_{22} + \Gamma_{31}\rho_{33} + \Gamma_{41}\rho_{44}, \\
\dot{\rho}_{22} &= i\Omega_p^*\rho_{32} - i\Omega_p\rho_{23} + i\Omega_d^*\rho_{42} - i\Omega_d\rho_{24} - \Gamma_{21}\rho_{22} + \Gamma_{32}\rho_{33} + \Gamma_{42}\rho_{44}, \\
\dot{\rho}_{33} &= -i\Omega_c^*\rho_{31} + i\Omega_c\rho_{13} - i\Omega_p^*\rho_{32} + i\Omega_p\rho_{23} - (\Gamma_{31} + \Gamma_{32})\rho_{33} + \Gamma_{43}\rho_{44}, \\
\dot{\rho}_{21} &= -[i(\Delta_c - \Delta_p) + \gamma_{21}]\rho_{12} - i\Omega_c\rho_{23} + i\Omega_p^*\rho_{31} + i\Omega_d^*\rho_{41}, \\
\dot{\rho}_{31} &= -(i\Delta_c + \gamma_{31})\rho_{31} + i\Omega_c\rho_{11} - i\Omega_c\rho_{33}, \\
\dot{\rho}_{41} &= -[i(\Delta_c + \Delta_d - \Delta_p) - \gamma_{41}]\rho_{41} + i\Omega_d\rho_{21} - i\Omega_c\rho_{43}, \\
\dot{\rho}_{32} &= -(i\Delta_p + \gamma_{32})\rho_{23} + i\Omega_c\rho_{12} - i\Omega_d\rho_{34} + i\Omega_p\rho_{22} - i\Omega_p\rho_{33}, \\
\dot{\rho}_{42} &= -(i\Delta_d + \gamma_{42})\rho_{42} + i\Omega_d\rho_{22} - i\Omega_d\rho_{44}, \\
\dot{\rho}_{43} &= [i(\Delta_p - \Delta_d) - \gamma_{43}]\rho_{43} - i\Omega_c^*\rho_{41} + i\Omega_d\rho_{23} - i\Omega_p^*\rho_{42}.
\end{aligned} \quad (2)$$

The above equations are restricted by the conjugation condition $\rho_{ij} = \rho_{ji}^*$ and the trace condition $\sum_i \rho_{ii} = 1$. Where the complex coherence dephasing rate on the transition $|i\rangle$ to $|j\rangle$ is denoted by $\gamma_{ij} = (\Gamma_i + \Gamma_j)/2$. Here, the relevant population decay rates are $\Gamma_i = \Sigma_k \Gamma_{ik}$ and $\Gamma_j = \Sigma_k \Gamma_{jk}$, with $k = 1, 2, 3$ and $4$ describing the inevitable dissipation within the system, and $\Gamma_{31} = \Gamma_{32} = \Gamma_{41} = \Gamma_{42} = \Gamma$. Under the steady state $\dot{\rho}_{ij} = 0$, we can obtain the numerical solution of $\rho_{32}$, which is governed by the probe detuning $\Delta_p$. Then, the steady state probe susceptibility $\chi_p$ of each filled cell can be obtained in $N$-type atomic system, given by

$$\chi_p(x) = \frac{N(x) |\mathbf{d}_{23}|^2 \rho_{32}}{\hbar \varepsilon_0 \Omega_p}. \quad (3)$$

Here, the real and imaginary parts of susceptibility, corresponding to dispersion and absorption lines of probe beam, $\varepsilon_0$ is the dielectric constant in vacuum, and the spatial atomic density $N(x)$, which can be considered as a Gaussian distribution in each filled cell, is given by

$$N(x) = \frac{N_0 \lambda_0}{\sigma_x \sqrt{2\pi}} \cdot e^{[-(x-x_0)^2/2\sigma_x^2]}. \quad (4)$$

Where $N_0$ is the average atomic density, $x_0$ is the trap center, $\sigma_x = \lambda_{Lat}/(2\pi\sqrt{\eta})$ is the half-width with $\eta = 2U_0/(\kappa_B T)$ related to the capture depth $U_0$ and temperature $T$, and $\lambda_0 = \lambda_{Lat}/2$ is the width of each cell, with $\lambda_{Lat}$ being the wavelength of a red-detuned retroreflected laser beam forming the optical lattice. For the Bragg condition, the sufficiently small incident angle is given by $\theta = \arccos(\lambda_p/\lambda_{Lat0})$, where $\lambda_{Lat0} = \lambda_{Lat} - \triangle\lambda_{Lat}$ with the geometric Bragg shift $\triangle\lambda_{Lat}$. It is worth emphasizing that the condition for trapping atoms is $\lambda_{Lat} > \lambda_{32}$, where $\lambda_{32}$ is the wavelength of the transition $|3\rangle \leftrightarrow |2\rangle$. Under $\lambda_p \approx \lambda_{Lat}$, the probe light can undergo efficient nonlinear Bragg scattering in the defective atomic lattice.

The reflection and transmission properties of the defective lattice can be characterized by a $2 \times 2$ unimodular transfer matrix. For filled lattice cells, we divide each cell into enough thin layers that can be considered homogeneous media. Generally, the reflection coefficient complexly depends on the layer thickness for layers of finite size. However, the transfer matrix method is valid in one-dimensional atomic lattice, as the transfer matrix for a vacuum layer is unity. With the transfer matrix $m(x_j)$ of a single layer, we can obtain reflection coefficient $r^l(x_j) = -m(2,1)/m(2,2)$ $[r^r(x_j) = m(1,2)/m(2,2)]$ on the left (right) side of this layer($m$ is the thin-layer transmission matrix, $l$ and $r$ represent different incident directions), depending on the Fresnel coefficients which are determined by the refractive index $n_f(x) = \sqrt{1 + \chi_p(x)}$. For the homogeneous and thin layer $r^l(x_j) = r^r(x_j) \equiv r(x_j)$, then $m(x_j)$ can be written as

$$m(x_j) = \frac{1}{t(x_j)} \begin{bmatrix} t(x_j)^2 - r(x_j)^2 & r(x_j) \\ -r(x_j) & 1 \end{bmatrix}. \quad (5)$$

Here, $t(x_j)$ and $r(x_j)$ are the transmission and reflection coefficients of $j$th layer in each lattice cell with $j \in [1, 100]$. The transfer matrix of one filled cell can be expressed as

$$M_f = \Pi_{j=1}^{100} m(x_j). \quad (6)$$

The refractive index $n_v = 1$ for vacant lattice cells. Thus, the transfer matrix of a vacant cell is written as

$$M_v = \frac{1}{t(x)} \begin{bmatrix} t(x)^2 & 0 \\ 0 & 1 \end{bmatrix} = \begin{bmatrix} e^{ik\lambda_0} & 0 \\ 0 & e^{-ik\lambda_0} \end{bmatrix}. \quad (7)$$

Then, the total transfer matrices is represented as

$$M_d = (M_f)^{P_2} \cdot [(M_v)^{a_v} \cdot (M_f)^{a_f}]^{P_1}. \quad (8)$$

The reflectivity on both sides of the 1D defective atomic

lattice can be obtained:

$$R^l = |r^l|^2 = \left|-\frac{M(2,1)}{M(2,2)}\right|^2,$$
$$R^r = |r^r|^2 = \left|\frac{M(1,2)}{M(2,2)}\right|^2. \quad (9)$$

Similarly, the $S$ matrix relates the outgoing (electric) field amplitudes $E^r_{pr}$ and $E^r_{pl}$ to the incoming (electric) field amplitudes $E^i_{pl}$ and $E^i_{pr}$ [see Fig. 1(c)]

$$\begin{bmatrix} E^r_{pr} \\ E^r_{pl} \end{bmatrix} = S \begin{bmatrix} E^i_{pl} \\ E^i_{pr} \end{bmatrix} = \begin{bmatrix} t & r^r \\ r^l & t \end{bmatrix} \begin{bmatrix} E^i_{pl} \\ E^i_{pr} \end{bmatrix}, \quad (10)$$

to verify the existence of lasing, we use the eigenvalues $\lambda_\pm$ of $S$ matrix to explore the spectral singularities (SS):

$$\lambda_\pm = t \pm \sqrt{r^l r^r}, \quad (11)$$

where we can identify SS through $\lambda_+ \to \infty$. To make it easier to identify, we use the inverse of $S$ matrix, and its eigenvalues $\lambda_\pm^{-1}$ are expressed in terms of the elements of the transfer matrix [61], namely

$$\lambda_\pm^{-1} = \frac{M(2,2)}{1 \pm \sqrt{1 - M(1,1)M(2,2)}}. \quad (12)$$

Besides, to investigate the origin of nonreciprocal photonic bandgaps, we explore the Bragg conditions for the entire defective atomic lattice:

$$\frac{\omega_p}{c}\left[p_1(a_f\lambda_0\overline{n_f(x)} + a_v\lambda_0 n_v) + p_2\lambda_0\overline{n_f(x)}\right] = \mathbb{N}\pi, \quad (13)$$

where $c$ is the speed of light in vacuum, $\mathbb{N}$ is an arbitrary positive integer, and $\overline{n_f(x)}$ is the average refractive index of a filled cell. In the defective atomic lattice, Eq. (13) can be used to analyze the conditions required for the appearance of photonic bandgaps. However, for more complex cases of unidirectional light amplification, the unidirectional photonic bandgaps may emerge, for which this Bragg condition has certain limitations. The more specific expression of valley-point will be provided later.

## III. NUMERICAL RESULTS AND DISCUSSIONS

In this section, we have two primary objectives: I) to discuss the conditions and physical essence for achieving URL; II) to explore the approaches to regulate and control URL. Firstly, we examine the probe reflectivities on both sides $v.s.$ $a_f$ and $\Delta_p$, with assuming $a_v = 10 \cdot a_f$ and other parameters are determined. As shown in Fig. 2(a), the $\log_{10} R^l$ and $\log_{10} R^r$ exhibit same tendency (increase or decrease with $a_v$), when $a_f = 10$ and $a_f = 20$. However, we find that the right reflection exhibits a valley around probe detuning $\Delta_p = -20$ MHz with $a_f = 15$, which corresponds to the peak of the left reflection; we define this point as the valley-point. It is evident that the emergence of the valley-point is accompanied by a decrease in the right reflection. According to our previous studies [56, 57], the introduction of the vacant lattices breaks the spatial symmetry of the susceptibility, leading to pronounced optical nonreciprocity. In particular, the period $P_1$ of Part I plays a crucial role in controlling both the optical nonreciprocity and the emergence of the valley-point. Basis on Fig. 1(a), to further investigate the modulation of defective region in the atomic lattice (Part I) on the valley-point, we plot the left and right reflectivity as functions of $P_1$ and $\Delta_p$ [see Fig. 2(b$_1$) and 2(b$_2$)]. We can clearly see that the valley-point of $R^r$ exhibits a periodic occurrence with the period of 5, and correspondingly, a significant enhancement of the left reflection peak. Therefore, we set $a_f = 15$ and $P_1 = 41$ in our regime, to balance high amplification with a small-scale atomic lattice. Once the lattice structure is fixed, modulating the external field to control the propagation characteristics of the probe light represents a simpler and more direct approach. Subsequently, when we slightly reduce the intensities of the two coupling fields, we find that the valley-point of the right reflection not only decreases in magnitude but also shifts toward the peak of the left reflection, and corresponding to the significantly increased left reflection. It is worth emphasizing that the valley-point of right reflection decreased from 14.7 to 2.6, and the peak of left reflection increased from $4.2 \times 10^3$ to $2 \times 10^4$ [see Fig. 2(c) and 2(d)]. In further, by finely adjusting the intensity of the strong coupling fields to $\Omega_c = 36.5$ MHz and $\Omega_d = 18.5$ MHz (purple lines), as shown in the insets of Fig. 2(c) and 2(d). The valley-point of $R^r$ has decreased to 0.25 while the peak of corresponding $R^l$ has reached to $1.2 \times 10^5$, at $\Delta_p = -18.0$ MHz. It is not difficult to observe that the lower the valley-point of the right reflection, the better the unidirectional amplification of the left reflection. Can the URL be achieved by modulating the valley-point? What is the physical essence underlying the generation of the valley-point? These questions have motivated our research interest in it.

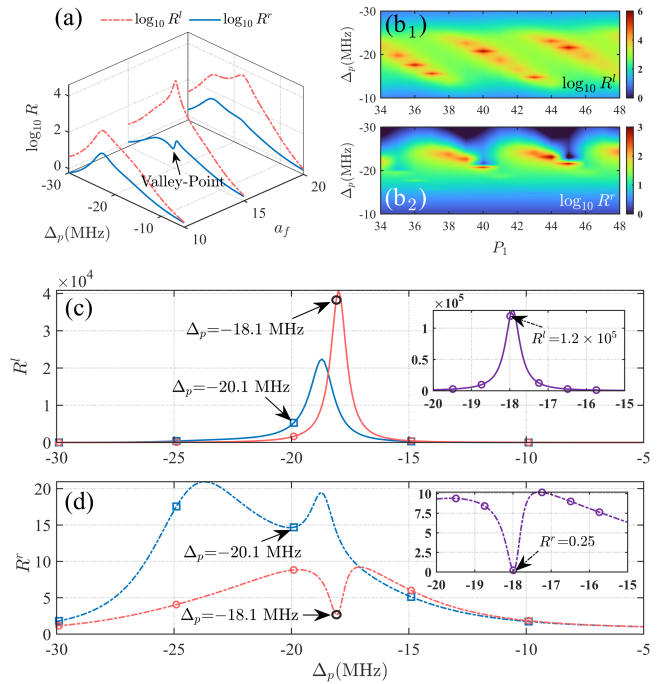

FIG. 2: (a) The logarithm of the probe reflectivities on both sides, $\log_{10} R^l$ and $\log_{10} R^r$, versus probe detuning $\Delta_p$ and $a_f$ of Part I with $a_v = 10 \cdot a_f$ in $\Omega_c=40$ MHz, $\Omega_d=20$ MHz, $P_1=35$. (b$_1$)&(b$_2$) The $\log_{10} R^l$ and $\log_{10} R^r$ versus $P_1$ periods of Part I with $\Omega_c=40$ MHz & $\Omega_d=20$ MHz in $a_f=15$. (c)&(d) The probe reflectivities of both sides, $R^l$ (solid line) and $R^r$ (dash-dotted line), versus $\Delta_p$ with $\Omega_c=40$ MHz & $\Omega_d=20$ MHz (blue rectangles), $\Omega_c=36$ MHz & $\Omega_d=18$ MHz (red circles), $\Omega_c=36.5$ MHz & $\Omega_d=18.5$ MHz (purple circles) respectively, with $a_f=15$, $P_1=41$. Here, $P_2=1200$, $\Delta_c=-15$ MHz, $\Delta_d=20$ MHz, and other relevant parameters for all panels are the same as Fig. 1.

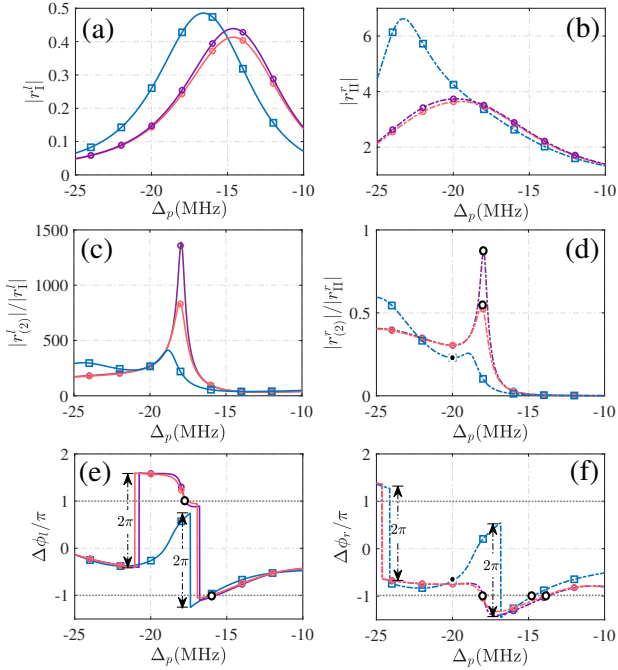

FIG. 3: (a)&(b) The modulus of probe direct reflection coefficients of both sides, $|r^l_{\rm I}|$ in Part I and $|r^r_{\rm II}|$ in Part II, versus probe detuning $\Delta_p$. (c)&(d) The secondary and direct reflection coefficients ratio $|r^l_{(2)}|/|r^l_{\rm I}|$ and $|r^r_{(2)}|/|r^r_{\rm II}|$ versus $\Delta_p$. (e)&(f) The relative phase $\Delta\phi_l$ and $\Delta\phi_r$ versus $\Delta_p$. The color scheme and line styles are identical to those in Fig. 3 and other parameters are the same as Fig. 1.

The reflectivity of the whole medium can be easily expressed using the reflection and transmission coefficients of Part I and Part II through transfer matrix, namely

$$R^l = |r^l|^2 = \left| r^l_{\rm I} + \frac{t^2_{\rm I} r^l_{\rm II}}{r^r_{\rm I} r^l_{\rm II} - 1} \right|^2,$$
$$R^r = |r^r|^2 = \left| r^r_{\rm II} + \frac{t^2_{\rm II} r^r_{\rm I}}{1 - r^r_{\rm I} r^l_{\rm II}} \right|^2. \quad (14)$$

Here, the left- and right-side reflectivities of Part I (II) are reciprocal, namely $|r^l_{\rm I}| = |r^r_{\rm I}|$ ($|r^l_{\rm II}| = |r^r_{\rm II}|$). The left-side reflection originates from the direct reflection in Part I and the secondary reflection in Part II of the transmitted wave from the Part I. Importantly, the secondary reflection is associated with the interaction term (i.e., $\widetilde{\mathcal{I}} := 1 - r^r_{\rm I} r^l_{\rm II}$) between the two parts. The right-side reflection follows similarly. Therefore, we define the secondary reflection of left- and right- side as

$$r^l_{(2)} = \frac{t^2_{\rm I} r^l_{\rm II}}{r^r_{\rm I} r^l_{\rm II} - 1}, \; r^r_{(2)} = \frac{t^2_{\rm II} r^r_{\rm I}}{1 - r^r_{\rm I} r^l_{\rm II}}. \quad (15)$$

To analyze the problem more clearly, we present the complex-valued expressions corresponding to the reflection coefficients, that Eq. (14) can be rewritten as

$$r^l = |r^l_{\rm I}| e^{i\phi_{l_{\rm I}}} \left( 1 + \frac{|r^l_{(2)}|}{|r^l_{\rm I}|} e^{i\Delta\phi_l} \right),$$
$$r^r = |r^r_{\rm II}| e^{i\phi_{r_{\rm II}}} \left( 1 + \frac{|r^r_{(2)}|}{|r^r_{\rm II}|} e^{i\Delta\phi_r} \right), \quad (16)$$

here $\Delta\phi_{l(r)}$ represent the relative phase between the direct reflection and the secondary reflection. It is not difficult to see from Eqs. (14) and (16) that the phase $\phi_{l_{\rm I}}(\phi_{r_{\rm II}})$ of the direct reflection does not affect the reflectivity. That is to say, the direct reflection of left or right reflection depends only on the reflection coefficients corresponding to Part I or Part II, while the secondary reflection on the left (right) side occurs after transmission through Part I (II), followed by reflection from Part II (I), and then transmission through Part I (II) again, which is modulated by the relative phase $\Delta\phi_l$ ($\Delta\phi_r$).

Next, we will conduct an in-depth analysis of how direct reflection and secondary reflection modulate the left and right reflectances. It is obviuous that the direct reflection in Part II is much larger than that in Part I, which can be observed from the modulus of the direct reflection coefficients in Figs. 3(a) and 3(b). The probe light can be amplified by the coherent gain atomic system; particularly, when passing through Part II, whose relatively large spatial length satisfies the non-linear Bragg scattering condition in periodically filled lattice cells, that the probe light can form an amplified reflection band modulated by the coupling fields. This ultimately results yielding $|r^r_{\rm II}| \gg |r^l_{\rm I}|$. However, the secondary reflection of Part I is much larger than that of Part II, exhibited as the ratios $|r^l_{(2)}|/|r^l_{\rm I}| \gg |r^r_{(2)}|/|r^r_{\rm II}|$ in Figs. 3(c) and 3(d), respectively. The total probe reflectivity after passing through Part I and Part II is the coherent superposition of direct reflection and secondary reflection, which depends on the relative phase $\Delta\phi_{l(r)}$. When $\Delta\phi_{l(r)} = (2\mathbb{N}+1)\pi$, $\mathbb{N}$ is an integer, accompanied by destructive interference, which well corresponds to the right reflection [see Fig. 3(f)]. However, destructive interference cannot be achieved in the left reflection [see Fig. 3(e)]. These are completely consistent with the right-reflection valley and the left-reflection peak in Fig. 2. The fundamental reason depends on the ratio $|r^r_{(2)}|/|r^r_{\text{II}}|$, the closer this ratio is to 1, the more obvious the destructive interference becomes. For the ratio $|r^l_{(2)}|/|r^l_{\text{I}}| \gg 1$, there is a significant difference between the direct reflection and the secondary reflection, thereby preventing the occurrence of interference effects. It is not difficult to see that under different coupling field intensities, as the right-reflection valley moves toward the left-reflection peak in Figs. 2(c) and 2(d), correspondingly, the right-side ratio approaches 1, and the relative phase $\Delta\phi_r$ approaches $\pi$ in Figs. 3(e) and 3(f). In particular, when $\Delta\phi_r = -\pi$, it implies that a complete destructive interference occurs between the direct reflection and the secondary reflection on the right side. This explains why the right-reflection valley in Fig. 2(c) and 2(d), which almost perfectly corresponds to the left-reflection peak, nearly drops to zero ($R^r = 0.25$) can be called destructive interference point (DIP). Meanwhile, the left-side ratio $|r^l_{(2)}|/|r^l_{\text{I}}|$ also increases with the strengthening of right-side destructive interference.

Furthermore, we attempt to search for DIP close to zero by modulating the lattice structure and external fields, which is expected to realize high-efficiency unidirectional reflection amplification and even the URL. In Figs. 4(a) and 4(b), the right-reflection DIP depth is highly robust against variations in the period of Part II ($P_2$), even as $P_2$ increases. However, as $P_2$ increases, this DIP gradually narrows and may even disappear entirely. Correspondingly, the left-reflection peak rises with increasing $P_2$, eventually reaching its maximum value and stabilizing once $P_2$ reaches 1200. Then, with $P_2 = 1200$, we slightly modulated the coupling detuning $\Delta_c$ and observed that the reflection response is highly sensitive to $\Delta_c$. When $\Delta_c$ varies from $-15$ MHz to $-14$ MHz, the intensity of the left-reflection peak remains nearly unchanged, while the peak itself shifts rightward by approximately 1 MHz. The right-reflection DIP also shifts accordingly, and its valley value decreases by almost an order of magnitude, shown in Fig. 4(c) and 4(d).

Next, we investigate whether the DIP corresponds to URL. We focus on the DIP corresponding to $\Delta_p = -17.12$ MHz [red line in Fig. 4(d)]. In Fig. 5(a) and 5(b), it is found that at this point, the two eigenvalues of the inverse scattering matrix are equal to $M(2,2)$ and approach 0 ($\text{Re}[\lambda_\pm^{-1}] = \text{Re}[M(2,2)] = -1.8 \times 10^{-3}$ and $\text{Im}[\lambda_\pm^{-1}] = \text{Im}[M(2,2)] = 2.5 \times 10^{-2}$ at $\Delta_p = -17.12$ MHz), which exhibits the characteristics of spectral singularities (SS) for $M(2,2) \to 0$, indicating lasing output. Meanwhile, it also shows the feature of non-Hermitian degeneracy (NHD) for $\lambda_+^{-1} = \lambda_-^{-1}$, which implies the realization of unidirectional reflection (i.e., either the left or right reflection approaches 0). To further verify which of the left and right reflections approaches 0, we plot $M(2,1)$ and $M(1,2)$, which determine the left and right reflectivities respectively, as shown in Fig. 5(c). It is observed that at this point ($\Delta_p = -17.12$ MHz), $M(1,2) \to 0$ (i.e., $\text{Re}[M(1,2)] = 3.4 \times 10^{-3}$, $\text{Im}[M(1,2)] = 2.7 \times 10^{-4}$), meaning the right reflectivity approaches 0 [see Eq. (9)], which is in full correspondence with the DIP. It is worth emphasizing that under the constraint that $\det(M) = 1$, the occurrence of URL characterized by $M(2,2) \to 0$ and $M(1,2) \to 0$, implies that both $M(1,1)$ and $M(2,1)$ will diverge, which further indicates the convergence of $\lambda_-^{-1}$ to $\lambda_+^{-1}$. Notably, if $M(1,2)$ tends to 0 at a rate faster than $M(2,2)$, the right-reflection DIP emerges, with $\lambda_\pm^{-1} \to M(2,2)$. It is thus evident that the right-reflection DIP corresponds to the left-reflection lasing point, i.e., we can achieve the URL via destructive in-terference.

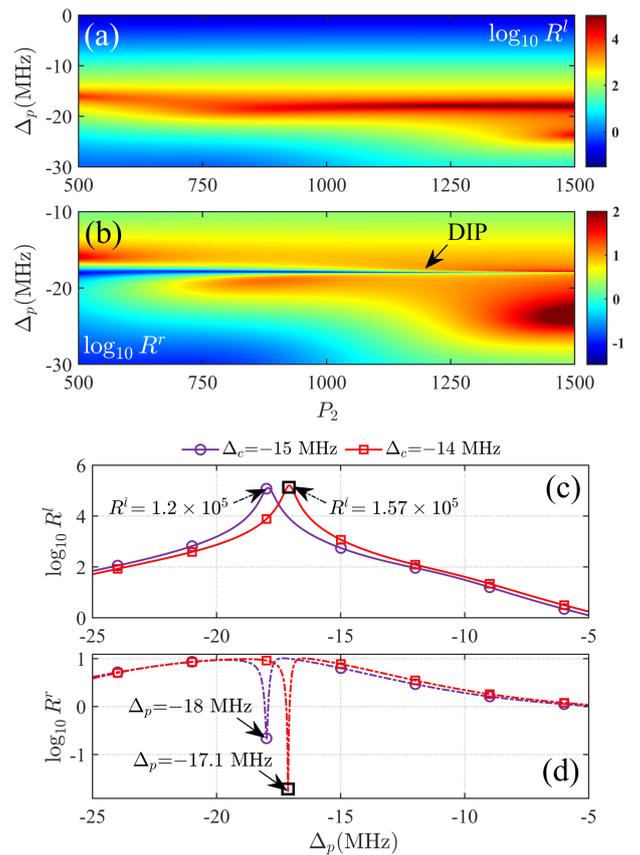

FIG. 4: (a)&(b) The $\log_{10} R^l$ and $\log_{10} R^r$ versus probe detuning $\Delta_p$ and $P_2$ periods of Part II, with $\Delta_c = -15$ MHz. (c)&(d) The $\log_{10} R^l$ and $\log_{10} R^r$ versus $\Delta_p$ in $P_2 = 1200$, with $\Delta_c = -15$ MHz (purple line) and $-14$ MHz (red line) respectively. Here, $\Omega_c = 36.5$ MHz, $\Omega_d = 18.5$ MHz, $\Delta_d = 20$ MHz, $a_f = 15$, $a_v = 150$, $P_1 = 41$, and other parameters are the same as Fig. 1.

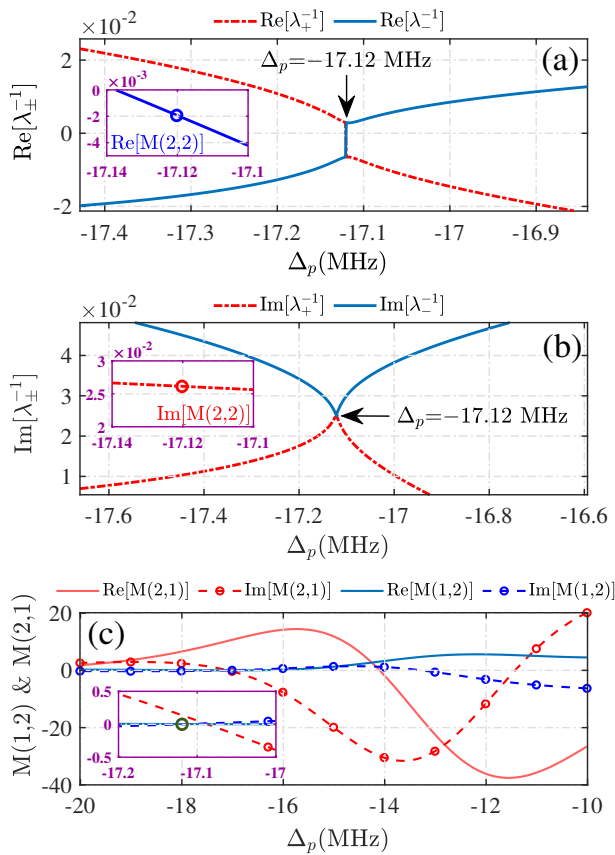

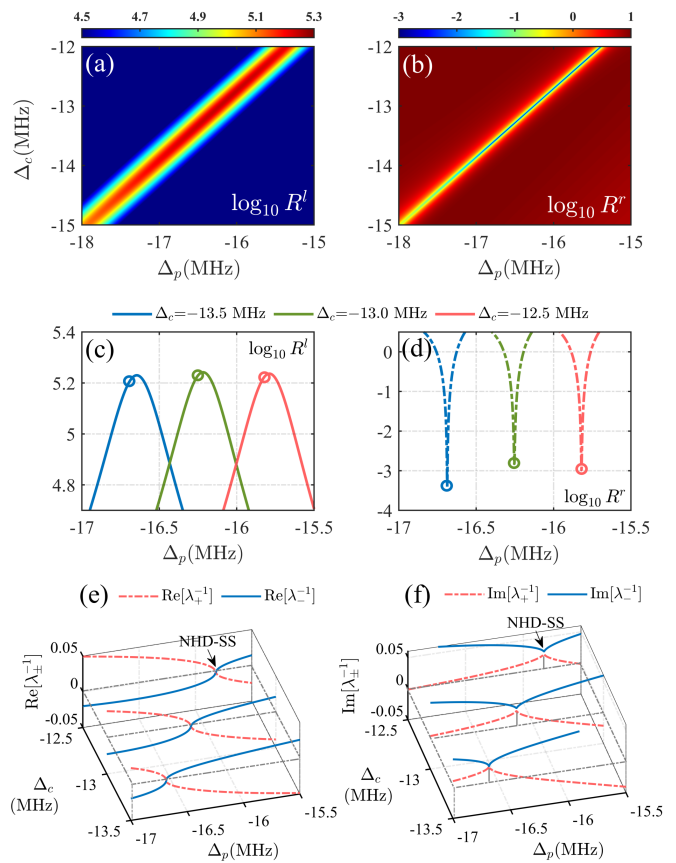

FIG. 5: (a)&(b) The real and imaginary part of eigenvalues $\lambda_\pm^{-1}$ of the inverse matrix of $S$ matrix versus probe detuning $\Delta_p$. (c) The non-diagonal elements $M(1,2)$ and $M(2,1)$ of transfer matrix versus $\Delta_p$. Here, $\Omega_c$=36.5 MHz, $\Omega_d$=18.5 MHz, $\Delta_c$=−14 MHz, $\Delta_d$ = 20 MHz, $a_f$=15, $a_v$=150, $P_1$=41, $P_2$=1200, and other parameters are the same as Fig. 1.

FIG. 6: (a)&(b) The $\log_{10} R^l$ and $\log_{10} R^r$, versus probe detuning $\Delta_p$ and coupling detuning $\Delta_c$. (c)&(d) Unidirectional reflection lasing modes under different $\Delta_c$. (e)&(f) The real and imaginary part of eigenvalues $\lambda_\pm^{-1}$ of inverse of $S$ matrix versus $\Delta_p$ and $\Delta_c$. Here, $\Omega_c$=36.5 MHz, $\Omega_d$=18.5 MHz, $\Delta_d$=20 MHz, $a_f$=15, $a_v$=150, $P_1$=41, $P_2$=1200. The other parameters are the same as Fig. 1.

TABLE I: The variation of $\log_{10} R$ and its corresponding probe frequency $\Delta_p$ under the modulation of coupling detuning $\Delta_c$, and other parameters are the same as Fig. 6(a)&(b).

| $\Delta_c$(MHz) | -14 | -13.5 | -13 | -12.5 | -12 | -11.5 |
|---|---|---|---|---|---|---|
| $\log_{10} R^r_{min}$ | -1.76 | -3.41 | -2.83 | -2.98 | -2.78 | -1.51 |
| $\Delta_p$(MHz) | -17.12 | -16.69 | -16.25 | -15.82 | -15.38 | -14.95 |
| $\log_{10} R^l$ | 5.18 | 5.21 | 5.23 | 5.23 | 5.21 | 5.17 |
| linewidth(kHz)[$R^r \leq 10^{-2}$] | – | 12 | 11 | 11 | 11 | – |

How to modulate the URL induced by destructive interference is worth to discussion in-depth. For the DIP is highly sensitive to $\Delta_c$, we plot the variation of left- and right- reflection with respect to $\Delta_c$ and $\Delta_p$. As clearly illustrated in Figs. 6(a) and 6(b), the left-reflection lasing undergoes a blue shift with the increase of $\Delta_c$, accompanied by a consistent trend of the right-reflection DIP. When $\Delta_c > -14$ MHz, the intensity of the left-reflection lasing stabilizes at $10^5$, while the right-reflection DIP drops to as low as $10^{-3}$. This is more clearly seen by the left- and right- reflection profiles corresponding to three distinct values of $\Delta_c$ in Figs. 6(c) and 6(d). Moreover, for the three sets of destructive interference induced URL cases, the two eigenvalues of the inverse scattering matrix satisfies the relation $\lambda_+^{-1} = \lambda_-^{-1} \to 0$, as shown in Figs. 6(e) and 6(f). Furthermore, the smaller the DIP is, the closer the two eigenvalues of the inverse scattering matrix are to zero. When DIP< $10^{-2}$, the left reflection can be regarded as an ideal unidirectional lasing. Then we provide detailed data in Table I under different $\Delta_c$, furthermore, perfect URL for four modes is demonstrated in the range of $\Delta_c \in [-13.5$ MHz$, -12$ MHz$]$. For these modes, the DIP can be as low as approximately $10^{-3}$, while the intensity of left-reflection lasing remains nearly stable at $10^5$, corresponding to an ultra-narrow lasing linewidth of 11 kHz $\sim$ 12 kHz. We note that the right-reflection DIP does not perfectly align with the left-reflection peak, with a frequency difference of about 0.04 MHz, but nearly identical in magnitude. The presence of such a frequency difference is a natural phenomenon.

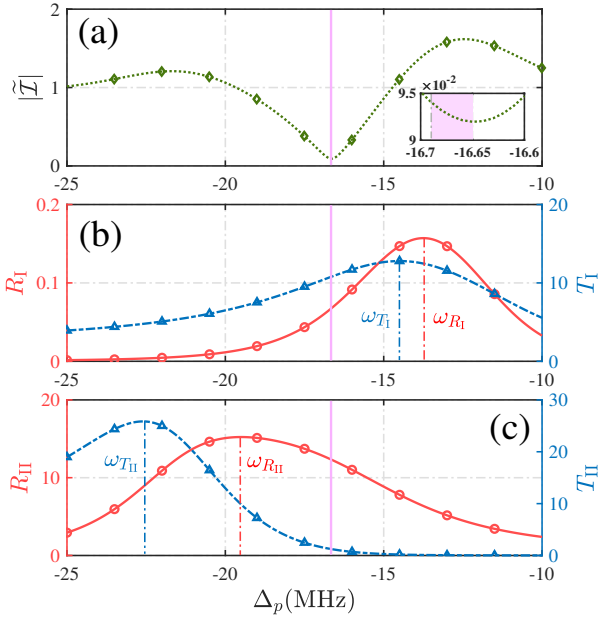

FIG. 7: (a) The modulus of interaction term $|\widetilde{\mathcal{I}}|$, (b) The reflectivity $R_\mathrm{I}$ and transmissivity $T_\mathrm{I}$ of Part I, (c) The $R_\mathrm{II}$ and $T_\mathrm{II}$ of Part II, respectively, versus $\Delta_p$. The parameters for all panels correspond to the unidirectional reflection lasing mode at $\Delta_c = -13.5$ MHz in Figs. 6(c) and 6(d).

As reported in our prior work [60], this discrepancy can be eliminated via the introduction and precise control of the optical field phase. However, such phase manipulation exerts a strong disruptive effect on the DIP. Next, we will discuss the origin of this frequency difference and the underlying mechanism by which the DIP induces reflection lasing.

Combining Eqs. (15) and (16) with Fig. 3, we conclude that the enhancement of left reflection originates from the ratio $|r^l_{(2)}|/|r^l_\mathrm{I}|$, while the attenuation of right reflection stems from destructive interference. As shown in Fig. 7(a), the interaction term $|\widetilde{\mathcal{I}}|$ reaches its minimum at $\Delta_p = -16.65$ MHz, which matches the left-reflection peak; by comparison, the right-reflection DIP lies at $\Delta_p = -16.69$ MHz, with a resultant corresponding frequency offset is 0.04 MHz (shaded in purple). More directly, two subsections (Part I and Part II) form a likewise Fabry-Pérot (FP) compound cavity governed by DFB mechanism. When the probe light cyclically oscillating between the two parts satisfies the standing wave condition $\phi_{r^l_\mathrm{II}}(\omega_p) + \phi_{r^r_\mathrm{I}}(\omega_p) = 2\mathbb{N}\pi$ and oscillation condition $|\widetilde{\mathcal{I}}(\omega_p)| \to 0$ at $\omega_p = \omega_L$, the inter-subsection feedback is strongly enhanced, yielding pronounced constructive interference and potentially reaching lasing threshold. As shown in Figs. 7(b) and 7(c), the DFB-dominated subsections possess distinct reflection-center frequencies $\omega_{R_\mathrm{I}}$ and $\omega_{R_\mathrm{II}}$. At their interface, an additional FP-like compound feedback channel arises, quantified by the interaction term $\widetilde{\mathcal{I}}$. The light incident from the left-side is not efficiently reflected in Part I (peak reflectivity is only 0.16 at $\omega_{R_\mathrm{I}}$); meanwhile, in a large probe frequency range, both the transmission band of Part I (centered at $\omega_{T_\mathrm{I}}$) and the reflection band of Part II (centered at $\omega_{R_\mathrm{II}}$) are strongly amplified by the probe gain. The entire process creates the light oscillation condition, thereby reaching lasing threshold and producing a greatly enhanced left-side secondary reflection $r^l_{(2)}(\omega_L)$, i.e., the left-reflection lasing peak. For right incidence, one component is directly reflected by Part II (centered at $\omega_{R_\mathrm{II}}$), while the other transmits through Part II, is reflected by Part I, and then transmits through Part II again. Since the transmission band of Part II (centered at $\omega_{T_\mathrm{II}}$) is far red-detuned from the reflection band of Part I (centered at $\omega_{R_\mathrm{I}}$), this pathway is essentially gain-free and leaves the right-side secondary reflection intrinsically extremely weak ($\ll 1$); thus, under the oscillation (threshold) condition, it is only amplified until it becomes comparable to the direct right reflection, i.e., $|r^r_{(2)}(\omega_L)|/|r^r_\mathrm{II}(\omega_L)| \to 1$, thereby laying the foundation for destructive interference in the right-reflection channel.

As illustrated in Fig. 3, the destructive interference between the right-side secondary reflection and the direct reflection is jointly determined by their relative amplitude and relative phase. When the right-incident probe light first propagates through the Part II and then undergoes reflection at the interface between two subsections, the corresponding total accumulated transmission phase is $2P_2\delta$. Herein, $\delta = \pi \cos\theta \sqrt{1+\overline{\chi}_p}$ denotes the propagation phase accumulated as probe light passes through one filled lattice cell, and $\delta \approx \pi$ for $|\overline{\chi}_p| \ll 1$. Consequently, the right-side reflection $R^r$ is governed by the relative phase difference $\Delta\phi'_r$ between $r^r_\mathrm{I}$ and $r^r_\mathrm{II}$. Once the standing wave and threshold conditions are satisfied at $\omega_p = \omega_L$, it follows that

$$\Delta\phi'_r = \phi_{r^r_\mathrm{I}} - \phi_{r^r_\mathrm{II}} = -(\phi_{r^l_\mathrm{II}} + \phi_{r^r_\mathrm{II}}) + 2\mathbb{N}\pi. \quad (17)$$

At $\omega_p = \omega_L$, the phase of the right-side reflection $R^r$ is determined entirely by Part II. According to the transfer matrix method based on Bloch theory, one obtains

$$\frac{r^l_\mathrm{II}}{r^r_\mathrm{II}} = -\frac{(M_f)^{P_2}(2,1)}{(M_f)^{P_2}(1,2)} = -\frac{M_f(2,1)}{M_f(1,2)} = -\frac{e^{-i\delta}-e^{i\delta}}{e^{i\delta}-e^{-i\delta}} = 1, \quad (18)$$

and from Eq. (18) it follows that $\phi_{r^l_\mathrm{II}} - \phi_{r^r_\mathrm{II}} = 2\mathbb{N}\pi$, then Eq. (17) can be rewritten as

$$\Delta\phi'_r = -2\phi_{r^l_\mathrm{II}} + 2\mathbb{N}\pi. \quad (19)$$

Only when $\Delta\phi'_r = (2\mathbb{N}+1)\pi$ (that is to say, $\phi_{r^l_\mathrm{II}} = \phi_{r^r_\mathrm{II}} = -\phi_{r^r_\mathrm{I}} = \pm\pi/2 + 2\mathbb{N}\pi$), the right-side reflection $R^r$ can be suppressed via destructive interference once $|r^r_{(2)}|/|r^r_\mathrm{II}| \to 1$, thereby producing a DIP, whereas the left-side reflection exhibits a pronounced lasing state. Otherwise, even the system satisfies the threshold condition, URL will not be generated; instead, nonreciprocal reflection lasing (NRL) will occur.

Overall, the emergence of URL (corresponding to the right-reflection DIP and the left-reflection lasing) is primarily determined by the following condition: (I) the reflection phase condition $\phi_{r_{\text{II}}^l} = \phi_{r_{\text{II}}^r} = -\phi_{r_{\text{I}}^r} = \pm\pi/2 + 2\mathbb{N}\pi$, which simultaneously satisfies the threshold, standing wave, and right-reflection destructive interference conditions. (II) the right-reflection amplitude condition $|r_{(2)}^r|/|r_{\text{II}}^r| = 1$, it governed by the near-threshold regime and attainable only due to threshold-induced feedback enhancement. Thus simultaneously satisfying the above two conditions necessarily results in the emergence of the right-reflection DIP and the corresponding URL. However, the right-reflection amplitude condition in general fails to coincide exactly with the lasing peak frequency $\omega_L$ and is most accurately satisfied at $\omega_V$, since it can be realized anywhere within the near-threshold regime (i.e., $\omega_V \to \omega_L$) and is not necessarily locked precisely at the defined threshold center (i.e., the lasing peak). Additionally, at $\omega_V$ slightly offset from $\omega_L$, the reflection phase condition deviates marginally, without appreciably weakening the destructive interference leading to the DIP. The resulting frequency mismatch $|\Delta\omega| = |\omega_L - \omega_V|$ is generally unavoidable, yet, this phenomenon can be eliminated via parameter fine-tuning. Briefly speaking, the right-reflection destructive interference arises from constructive feedback between the two subsections; consequently, an ideal right-reflection DIP is naturally accompanied by high-performance left-reflection lasing.

It is evident that investigating how lattice structure parameters modulate the DIP and how its stable existence can be sustained holds significant scientific value and practical importance. This is because the DIP serves as the core prerequisite for the generation of URL, implying that both the Bragg condition and the threshold condition are satisfied simultaneously, which corresponds to Eq. (13) and the relational expression proposed in our previous work [60], as

$$M_f(2,2) = \frac{e^{ika_v\lambda_0} - \widehat{P_2}\,\widehat{a_f}e^{-ika_v\lambda_0} - \widehat{P_2}\,\widehat{P_1}}{2i\widehat{P_2}\sin[ka_v\lambda_0]}. \quad (20)$$

The URL can be realized, when the matrix element $M_f(2,2)$ of one filled lattice satisfies Eq. (20). For example, using the parameters of Figs. 4(c) and 4(d) with $\Delta_c = -14$ MHz, $M_f(2,2) \approx -(1\pm\mathring{\eta})\pm i(\mathfrak{z}\pi/2 - \mathring{\iota})$, where $\mathring{\eta} \approx -2.0 \times 10^{-3}$ and $\mathring{\iota} \approx 2.2 \times 10^{-3}$ are both small quantities in the systems to satisfy the Bragg condition (i.e., "geometric" Bragg detuning $\mathfrak{z} = -2\Delta\lambda_{Lat}/\lambda_{Lat}$) and produce gain.

Then, based on Eq. (13), we utilize the properties of Chebyshev polynomials $U_n$ [62, 63] to expand $M_f(2,2)$ by replacing $\chi_p(x)$ with $\overline{\chi}_p$. Here, $U_n$ satisfies $U_n(\mathbb{A}\overline{\chi}_p + \mathbb{B}) \approx U_n(\mathbb{B}) + U_n'(\mathbb{B})\cdot\mathbb{A}\overline{\chi}_p$. Furthermore, we can obtain the more accurate Bragg condition of NHDSS that also characterizes the DIP in terms of $\mathfrak{z}$, $\mathring{\eta}$, and $\mathring{\iota}$, namely

$$\begin{aligned}\frac{\omega_p}{c}\lambda_0 \approx{}& \pi \pm \frac{1}{2+\mathfrak{z}}\cdot\Bigg\{\mathcal{C}_{P_2}\left(1 + \cot\left[\frac{\mathfrak{z}a_v\pi}{2+\mathfrak{z}}\right]\right)\\ &+ \mathcal{C}_{a_f}\left(1 - \cot\left[\frac{\mathfrak{z}a_v\pi}{2+\mathfrak{z}}\right]\right) \pm \alpha_v\mathscr{U}_1^{(P_1)}\\ &- 2 \mp 2\mathring{\iota} \mp \mathring{\eta}\left[2 + \alpha_v\mathscr{U}_2^{(P_1)}\right.\\ &\left.\pm\frac{\mathfrak{z}\pi}{2+\mathfrak{z}}\left(\tilde{\mathbb{C}}_1\cot\left[\frac{\mathfrak{z}a_v\pi}{2+\mathfrak{z}}\right] - \tilde{\mathbb{C}}_2\right)\right]\Bigg\},\end{aligned} \quad (21)$$

and $a_v = $ odd (even), $\alpha_v = -(+)\csc[\mathfrak{z}a_v\pi/(2+\mathfrak{z})]$, with

$$\begin{aligned}\mathcal{C}_{P_2} &= \frac{P_2}{P_2 - 1}, \; \mathcal{C}_{a_f} = \frac{a_f - 1}{a_f},\\ \tilde{\mathbb{C}}_1 &= \frac{\mathbb{C}_{P_2-1}^{(1)} - \mathcal{C}_{P_2}\mathbb{C}_{P_2-2}^{(1)}}{P_2 - 1} + \frac{\mathcal{C}_{a_f}\mathbb{C}_{a_f-1}^{(1)} - \mathbb{C}_{a_f-2}^{(1)}}{a_f},\\ \tilde{\mathbb{C}}_2 &= \frac{\mathbb{C}_{P_2-1}^{(1)} - \mathcal{C}_{P_2}\mathbb{C}_{P_2-2}^{(1)}}{P_2 - 1} - \frac{\mathcal{C}_{a_f}\mathbb{C}_{a_f-1}^{(1)} - \mathbb{C}_{a_f-2}^{(1)}}{a_f},\end{aligned} \quad (22)$$

and $\mathbb{C}_n^{(1)}$ denotes the coefficient of $\overline{\chi}_p^{(1)}$ in the Chebyshev polynomials. Herein, higher-order terms $\mathbb{C}_n^{(N)}$ and $\overline{\chi}_p^{(N)}$ (where $N \geq 2$) are neglected for simplification. Evidently, as $P_2$ increases to sufficiently large value, the coefficients $\tilde{\mathbb{C}}_1$ and $\tilde{\mathbb{C}}_2$ correlated with $P_2$ tend to stabilize. Consequently, the frequency $\omega_p$ corresponding to the DIP exhibits remarkable robustness over a specific range of $P_2$ as depicted in Figs. 4(a) and 4(b). Considering the complex internal embedded structure of $P_1$, $\mathscr{U}_1^{(P_1)}$ and $\mathscr{U}_2^{(P_1)}$ can be given

$$\begin{aligned}\mathscr{U}_{P_1}^{(1)} &= \frac{\left[\frac{\tilde{U}_{P_1-2}'}{\tilde{U}_{P_1-1}'} + \frac{\tilde{U}_{P_1-2}}{\tilde{U}_{P_1-1}}\right]\mathcal{Q}_{a_f}^{(1)}\left(\mathfrak{z} - \frac{2\mathring{\iota}}{\pi}\right) + \frac{\tilde{U}_{P_1-2}}{\tilde{U}_{P_1-1}'}}{a_f\left[\frac{\tilde{U}_{P_1-1}}{\tilde{U}_{P_1-1}'} + 2\mathcal{Q}_{a_f}^{(1)}\left(\mathfrak{z} - \frac{2\mathring{\iota}}{\pi}\right)\right]},\\ \mathscr{U}_{P_1}^{(2)} &= \frac{\left[\frac{\tilde{U}_{P_1-2}'}{\tilde{U}_{P_1-1}'} - \frac{\tilde{U}_{P_1-2}}{\tilde{U}_{P_1-1}}\right]\frac{2}{\pi}\mathcal{Q}_{a_f}^{(1)} - \frac{\tilde{U}_{P_1-2}}{\tilde{U}_{P_1-1}'}\frac{\mathfrak{z}\pi\mathbb{C}_{a_f-1}^{(1)}}{a_f(2+\mathfrak{z})}}{a_f\left[\frac{\tilde{U}_{P_1-1}}{\tilde{U}_{P_1-1}'} + 2\mathcal{Q}_{a_f}^{(1)}\left(\mathfrak{z} - \frac{2\mathring{\iota}}{\pi}\right)\right]},\end{aligned} \quad (23)$$

here $\tilde{U}_n$ represents the Chebyshev polynomial function of $\mathcal{Q}_{a_f}^{(2)}$, while $\tilde{U}_n'$ denotes its derivative with respect to $\mathcal{Q}_{a_f}^{(2)}$. Notably, $\mathcal{Q}_{a_f}^{(1)}$ and $\mathcal{Q}_{a_f}^{(2)}$ are the function of $a_f$, $a_v$, and $\mathfrak{z}$, given by

$$\begin{aligned}\mathcal{Q}_{a_f}^{(1)} &= \frac{\pi\alpha_{fv}}{2}\left(a_f - \frac{\mathfrak{z}\pi}{2+\mathfrak{z}}\mathscr{C}_{a_f}\cot\left[\frac{\mathfrak{z}a_v\pi}{2+\mathfrak{z}}\right]\right),\\ \mathcal{Q}_{a_f}^{(2)} &= -\alpha_{fv}\left(a_f\cdot\frac{\mathfrak{z}\pi}{2+\mathfrak{z}} + \cot\left[\frac{\mathfrak{z}a_v\pi}{2+\mathfrak{z}}\right]\right),\end{aligned} \quad (24)$$

where $\alpha_{fv} = \hat{\alpha}_f\hat{\alpha}_v\sin[\mathfrak{z}a_v\pi/(2+\mathfrak{z})]$ and $\mathscr{C}_{a_f} = \mathbb{C}_{a_f-1}^{(1)} - \mathbb{C}_{a_f-2}^{(1)} + a_f$. Here, $a_f = $ odd (even), $\hat{\alpha}_f = -1\,(+1)$, and $a_v = $ odd (even), $\hat{\alpha}_v = +1\,(-1)$. It follows that the parity of filled and vacant lattice cells dictates the position where the NHDSS emerge. Moreover, $P_1$ determines the periodic functional form of $\mathcal{Q}^{(2)}_{a_f}$ (i.e., $a_f$ and $a_v$), This further corroborates the aforementioned conclusion: as $P_1$ varies, the reflectivity $R^{l/r}$ exhibits a periodic oscillatory increase with a characteristic period that is inversely proportional to $\mathfrak{z}$ and not necessarily an integer. In Figs. 2(b$_1$) and 2(b$_2$), this regularity is reflected in $R^l$ peak and $R^r$ valley, with the period of $P_1$ being approximately 5. Clearly, the DIP exhibits a periodic dependence on $P_1$, with the period $T_{P_1}$ determined by $\mathcal{Q}^{(2)}_{a_f}$ and given by

$$T_{P_1} = \pi / \left\{ \frac{\pi - 2}{2} - \left( \frac{a_f}{a_v} - \frac{1}{2} \right) \cdot \left( \frac{\mathfrak{z} a_v \pi}{2 + \mathfrak{z}} \right)^2 \right\}. \quad (25)$$

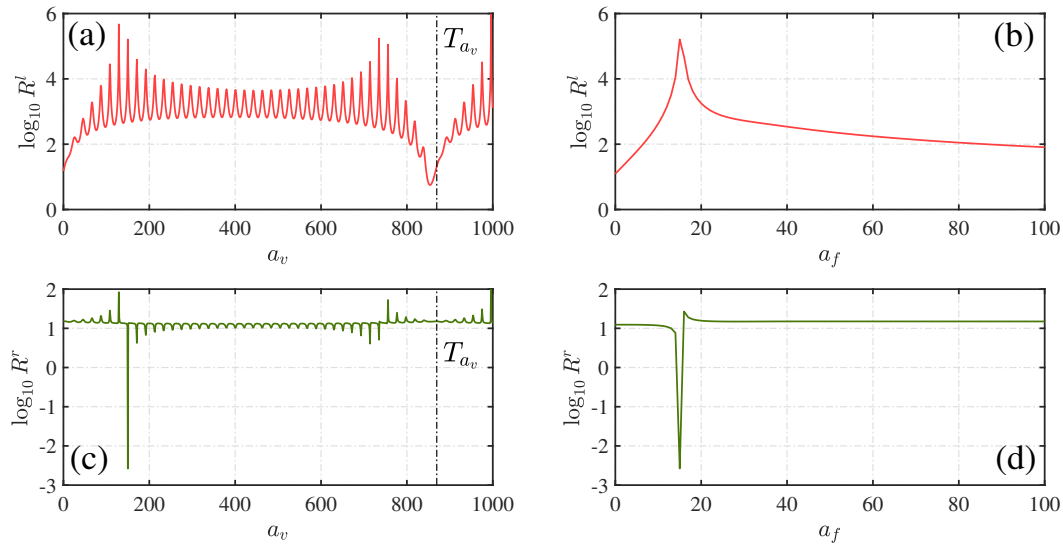

FIG. 8: (a)&(c) The $\log_{10} R^l$ and $\log_{10} R^r$ versus $a_v$ with $a_f$=15. (b)&(d) The $\log_{10} R^l$ and $\log_{10} R^r$ versus $a_f$ with $a_v$=150. Here, $\Omega_c$=36.5 MHz, $\Omega_d$=18.5 MHz, $\Delta_p$=−16.69 MHz, $\Delta_c$=−13.5 MHz, $\Delta_d$=20 MHz, $P_1$=41, $P_2$=1200. The other parameters for all panels are the same as Fig. 1.

Similarly, it can be seen from the combination of Eq. (21) and Eq. (24) that the parameter governing the Bragg condition, $\mathcal{Q}^{(2)}_{a_f}$, oscillates periodically with $a_v$, with a period given by

$$T_{a_v} = 1 + 2/\mathfrak{z}. \quad (26)$$

If $a_v = \mathbb{N} \cdot T_{a_v}$ ($\mathbb{N}$ is an arbitrary positive integer), the defective atomic lattice is equivalent to the perfect atomic lattice ($a_v = 0$), whereupon the left and right reflectivities are reciprocal. As illustrated in Figs. 8(a) and 8(c), when $a_v = 868$, the left reflectivity equals the right reflectivity ($\log_{10} R^l = \log_{10} R^r = 1.12$) after $P_1$ oscillation cycles, which corresponds to one oscillation cycle of $\mathcal{Q}^{(2)}_{a_f}$ ($\mathbb{N} = 1$, $a_v = T_{a_v}$). This reciprocal effect actually arises from the continuous accumulation of misalignment between the left-reflection peaks and right-reflection valleys within each small oscillation cycle. As observed herein, when $a_v = 150$, it corresponds to intense left-reflection lasing and a right-reflection DIP, which is also consistent with the case of $a_f = 15$ in Figs. 8(b) and 8(d). Furthermore, as $a_f$ increases, the left and right reflectivities exhibit strong robustness against $a_f$ and tend toward reciprocity. This conclusion can also be clearly derived from the relational expression in Eq. (22), when $a_f$ increases to a certain value, $\mathcal{C}_{a_f} \to 1$. In particular, the increase of $a_f$ also suppresses the spatial destructiveness of $a_v$, thereby reducing the nonreciprocity between the left and right reflectivities.

## IV. CONCLUSIONS

In summary, we propose a novel scheme for achieving narrow band URL (11 kHz) by leveraging the combined effects of a coherent gain atomic system (which provides probe gain) and a 1D defective atomic lattice [which introduces spatial symmetry breaking and a distributed feedback (DFB) regime], enabling the achievement of nonreciprocal reflection and lasing oscillation within a single system. In this system, there are two key factors for achieving URL: (I) The phase difference between the direct reflection and secondary reflection of the right reflection is an odd multiple of $\pi$, enabling destructive interference (corresponding to the right-reflection DIP). Meanwhile, the intensity of the secondary reflection of the left reflection is much higher than that of the primary reflection, which meets the threshold condition for lasing output (corresponding to left-reflection lasing). (II)

The lattice structure is appropriately modulated to ensure that the right-reflection DIP and left-reflection lasing satisfy the Bragg condition, thereby corresponding to the same frequency. Based on the above physical essence, numerical results and mathematical derivations, the relationships between the lattice structure parameters ($a_v$, $a_f$, $P_1$, $P_2$) and the left/right reflectivities are established. Accordingly, the structure of the defective atomic lattice can be precisely tailored to achieve the desired URL. In addition, the right-reflection DIP and the left-reflection lasing can be modulated by $\Omega_c$ and $\Omega_d$. Therefore, our work paves promising avenues for the realization of nonreciprocal optical circuits and high-performance all-optical controlled unidirectional devices tailored for integrated photonic systems.

## V. ACKNOWLEDGMENTS

This work is supported by the National Natural Science Foundation of China (Grant Nos. 12204137, 12564048, 12564047, 11874004, 11204019). This project is also supported by the Hainan Provincial Graduate Innovative Research Project (Grant No. Qhys2024-402).